\documentclass[twocolumn,amsmath,amssymb,pra]{revtex4}
\usepackage{graphicx}
\usepackage{dcolumn}
\usepackage{bm}

\begin{document}

\title{Sequential Quantum Teleportation of Optical Coherent States}

\author
{Hidehiro Yonezawa} \affiliation{Department of Applied Physics,
School of Engineering, The University of Tokyo, 7-3-1 Hongo,
Bunkyo-ku, Tokyo 113-8656, Japan} \affiliation{CREST, Japan
Science and Technology (JST) Agency, 1-9-9 Yaesu, Chuo-ku, Tokyo
103-0028, Japan}
\author{Peter van Loock}
\affiliation{National Institute of Informatics, 2-1-2
Hitotsubashi, Chiyoda-ku, Tokyo 101-8430, Japan}
\author
{Akira Furusawa} \affiliation{Department of Applied Physics,
School of Engineering, The University of Tokyo, 7-3-1 Hongo,
Bunkyo-ku, Tokyo 113-8656, Japan} \affiliation{CREST, Japan
Science and Technology (JST) Agency, 1-9-9 Yaesu, Chuo-ku, Tokyo
103-0028, Japan}


\begin{abstract}
We demonstrate a sequence of two quantum teleportations of optical
coherent states, combining two high-fidelity teleporters for
continuous variables. In our experiment, the individual
teleportation fidelities are evaluated as $F_{1}=0.70\pm0.02$ and
$F_{2}=0.75\pm0.02$, while the fidelity between the input and the
sequentially teleported states is determined as
$F^{(2)}=0.57\pm0.02$. This still exceeds the optimal fidelity of
one half for classical teleportation of arbitrary coherent states
and almost attains the value of the first (unsequential) quantum
teleportation experiment with optical coherent states.
\end{abstract}

\maketitle

\section{Introduction}

By utilizing shared entanglement and classical communication,
quantum teleportation \cite{Bennett93} enables one, in principle,
to transfer arbitrary quantum states with unit fidelity. In a
realistic scenario, typically, a receiver obtains an imperfect
version of the sender's state. If the receiver decides to teleport
his approximate version to a third party, the original quantum
state will further degrade when it arrives at the final
destination. Such a sequential quantum teleportation therefore
requires a sufficiently good performance of each individual
teleporter; otherwise the finally teleported state would hardly
resemble the input state.

Most protocols for quantum information processing and their
experimental realizations are based upon either discrete
qubit/qudit or continuous phase-space variables. In quantum
optical implementations, typically, the single-photon-based qubit
approach suffers from rather low efficiencies, but achieves, in
principle, near-unit fidelities. Combining several single-photon
teleporters \cite{Bouwmeester97,Boschi98} would, in principle,
still result in very good fidelities, though conditioned upon
coinciding detection events at very low success rates.

Conversely, when Gaussian resource states and continuous-variable
homodyne measurements are used, {\it unconditional operations lead
to, in principle, near-unit efficiencies, even when basic
subroutines such as quantum teleportation are concatenated}. The
continuous-variable approach
\cite{Braunstein_QICV(2003),Braunstein2005,EisertPlenio05},
however, as it relies on intrinsically imperfect squeezed-state
entanglement, will never result in arbitrarily high fidelities. In
continuous-variable (CV) quantum teleportation
\cite{Braunstein98}, the teleported state is a noisy replica of
the input state, with an excess noise depending on the quality of
the squeezed entanglement resource. Therefore, when quantum
information is sequentially manipulated through Gaussian
resources, for instance, via a sequence of teleportation circuits
in a continuous-variable cluster computation \cite{Menicucci06},
the unwanted excess noise accumulates and leads to increasingly
deteriorating fidelities. In order to achieve still better than
classical fidelities, it is thus crucial to improve the quality of
each individual teleporter.

So far, a variety of (unsequential) quantum teleportation
protocols have been demonstrated, for instance, with photonic
qubits \cite{Bouwmeester97,Boschi98}, optical field modes
\cite{Furusawa98}, between atoms \cite{Riebe04,Barrett04}, and
even between light and atoms \cite{Sherson06}. Since the first
realization of CV quantum teleportation of optical coherent states
\cite{Furusawa98}, several related experiments have followed
\cite{Sherson06,Bowen03a,Zhang03,Takei05e,Yonezawa04,Takei05s}.
The CV quantum teleporter can be characterized by the fidelity
$F=\left \langle \psi \right| \hat \rho_{out} \left| \psi \right
\rangle$ for an input state $\left| \psi \right \rangle$ and a
teleported state $\hat \rho_{out}$. If the input state is a
coherent state, $\left| \psi \right \rangle\equiv \left| \alpha
\right \rangle$, the total fidelity for a sequence of $n$ quantum
teleportations may be described by \cite{Suzuki06}
\begin{align}
F^{(n)}=1/ \left( 1+ne^{-2r} \right),
\end{align}
where $r$ is the squeezing parameter of the (equally) entangled,
standard two-mode squeezed-state resources. In the case of $n=2$,
at least $r=0.35$ is required (corresponding to two $F=2/3$
teleporters) in order to surpass the classical limit $F_{cl}=1/2$
\cite{Braunstein00,Braunstein01,Hammerer05}.

In the present work, we demonstrate an experiment of two
sequential quantum teleportations of optical coherent states. The
input states are teleported from a sender (``Alice 1'') to a first
receiver (``Bob 1'') and could be retrieved there, as we verify
through fidelity measurements. In a second round of teleportation,
the output states of the first teleporter are then transferred
from Bob 1 (now acting as ``Alice 2'') to a second receiver (``Bob
2'') where they can be verified again via fidelity measurements.
As we use two high-fidelity teleporters, the total fidelity of the
output states is still well beyond the classical limit, despite
the excess noise accumulated during the two rounds of quantum
teleportation. During the entire protocol, none of the
participants (the Alice's and Bob's) gain any substantial
information about the input state, though they do obtain some
partial knowledge because of the non-maximal degree of
entanglement of the finitely squeezed states used for
teleportation.

Based on these results, one could sequentially communicate
coherent signal states over two different segments of a channel,
provided the corresponding amount of entanglement is available in
each channel segment. For realistic, noisy channels, this would
require entanglement distillation (purification) procedures
\cite{Bennett96,Browne03} performed prior to the teleportations.
In general, however, by dividing a communication channel into
shorter segments \cite{Briegel98}, a higher degree of entanglement
can be maintained in each segment and entanglement distillation
will be more efficient. Via sequential quantum teleportation,
quantum information can then be sent to an intermediate station
where it could be either retrieved or passed on to the next
station.

If the total communication channel covers a large distance (e.g.,
1000 km), it will be necessary to connect some of the purified
entangled states via entanglement swapping \cite{Zukowski93} and
repurify the resulting states \cite{Briegel98}. Using various
levels of purification and swapping enables one to cover a larger
distance compared to sequential quantum teleportation. However,
after entanglement swapping, quantum information can no longer be
transferred to the intermediate stations and potentially retrieved
there. As a consequence, in CV entanglement swapping
\cite{Takei05e}, resource requirements are less demanding and, in
principle, any nonzero squeezing value of the initial two-mode
squeezed states results in a swapped entangled state sufficient
for $F>F_{cl}=1/2$ quantum teleportation of coherent states
\cite{vanloock00} (as opposed to the $r>0.35$ squeezing limit in
sequential quantum teleportation). A comparison between these
related schemes is shown in Fig.~1.

Concerning quantum computation rather than communication,
measurement-based schemes
\cite{Nielsen06,RaussendorfBriegel01,GottesmanChuang99} have been
shown to be an interesting alternative to the more traditional
circuit-based approach. The present experiment is also a first
step towards a sequential, measurement-based manipulation of
quantum information during its propagation through an efficient
Gaussian resource state \cite{Menicucci06,vanloock07}. A
comparison of the current sequential teleportation scheme with a
protocol in which an input state is sent through a linear Gaussian
cluster state is shown in Fig.~1.

 \begin{figure}[t]
     \begin{center}
        \includegraphics[width=1.0\linewidth]{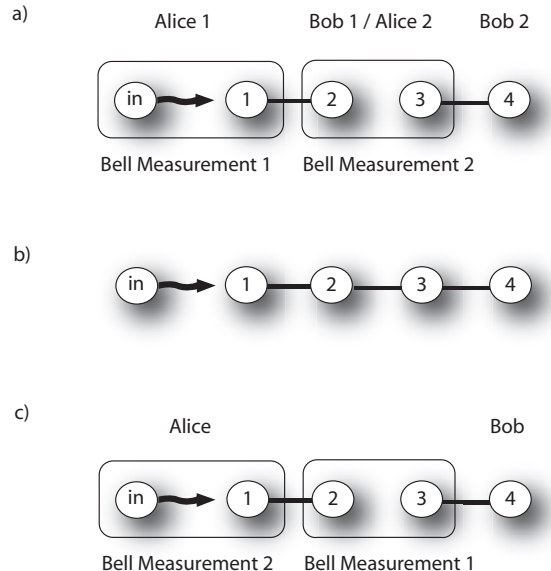}
        \caption{Comparison of sequential quantum teleportation (a)
        with the propagation of a quantum state through a linear cluster state (b) and
        entanglement swapping plus quantum teleportation (c).
        In the present experiment, the input coherent state is first teleported onto mode 2
        as in (a); after this first quantum teleportation, the input state, now present in mode
        2, could be verified by measuring its fidelity; the state of mode 2 is then again teleported,
        this time onto mode 4.
        In the cluster scenario (b), the input state is attached to the cluster
        and, in order to send it along the chain to mode 4, quadrature measurements
        are performed on each individual mode except mode 4; the entangling
        gate, attaching the input mode to the cluster, and the subsequent single-mode measurements of
        the input mode and mode 1 correspond to Bell measurement 1 in (a); however, different
        from (a), in (b), there is no intermediate occurrence of the teleported state at mode 2,
        as modes 2 and 3 have been entangled {\it prior} to any measurements; in
        (a), the entangling gate between 2 and 3 is postponed until Bell measurement 2.
        Another related scheme is (c), including entanglement swapping; here the first Bell measurement
        is performed on modes 2 and 3, leaving modes 1 and 4 in an entangled state (whose particular
        form depends on the measurement outcomes); finally the input state can be teleported onto mode 4
        via a second Bell measurement of the input mode and mode 1; again different from (a), in (c),
        there is no intermediate occurrence of the input state at mode 2.}
     \end{center}
   \end{figure}

\section{Sequential Quantum Teleportation Protocol}

The quantum state to be teleported here is that of an
electromagnetic field mode. An electromagnetic field mode is
represented by an annihilation operator $\hat{a}$ with real and
imaginary parts $\hat{x}$ and $\hat{p}$ corresponding to the
``position'' and ``momentum'' quadrature-phase amplitude
operators. These operators $\hat{x}$ and $\hat{p}$ satisfy the
commutation relation $[\hat{x},\hat{p}] =i/2$ (units-free, with
$\hbar =1/2$).

In our experiment, the input state is a coherent state for an
optical sideband at 1 MHz. The experimental setup is shown in
Fig.~2. In order to generate squeezed vacuum states, we use four
subthreshold optical parametric oscillators (OPOs) with a
periodically poled KTiOPO$_4$ as a nonlinear medium
\cite{Suzuki06}. An output of CW Ti:Sapphire laser at 860nm is
frequency doubled in an external cavity with a potassium niobate
crystal. The output beam at 430nm is divided into four beams to
pump the four OPOs. The pump powers are about 90mW. By combining
two squeezed-vacuum states at a symmetric beam splitter, we can
generate an entangled two-mode squeezed (``EPR'') state. Using
four squeezed vacuum states, we generate two pairs of EPR beams in
order to construct two teleporters.

   \begin{figure}[t]
     \begin{center}
        \includegraphics[width=1.0\linewidth]{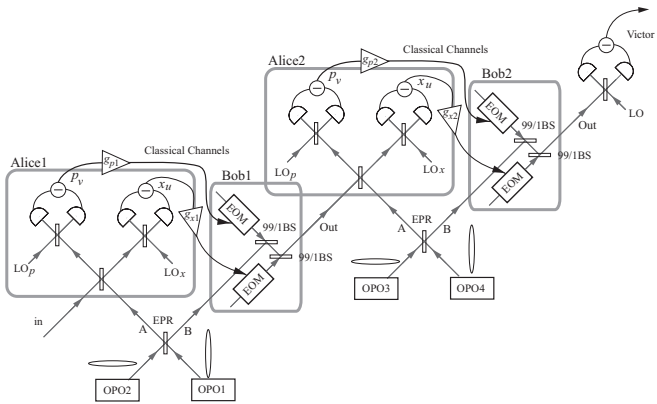}
        \caption{Experimental setup for sequential quantum teleportation.
        OPOs are optical parametric oscillators.
        EOMs are electro-optical modulators.
        All beam splitters except 99/1 BSs are 50/50 beam splitters.
        LOs are local oscillators for homodyne detection.}
     \end{center}
   \end{figure}

In the following, we describe the teleportation process in the
Heisenberg representation. Initially, the sender Alice and the
receiver Bob share a pair of EPR beams. Alice performs a joint
measurement on her EPR mode ($\hat x_{\rm A}$, $\hat p_{\rm A}$)
and the input mode ($\hat x_{in}$, $\hat p_{in}$). She combines
these two modes at a symmetric beam splitter and measures $\hat
x_u=\left( \hat x_{in} - \hat x_{\rm A} \right)/\sqrt{2}$ and
$\hat p_v=\left( \hat p_{in} + \hat p_{\rm A} \right)/\sqrt{2}$
with two homodyne detectors. The measurement results ($x_u$,
$p_v$) are then sent to Bob through classical channels with gain
$g_x$ and $g_p$.

The (normalized) gains of the classical channels are adjusted
similar to Ref. \cite{Zhang03} and defined as $g_x=\langle \hat
x_{out} \rangle / \langle \hat x_{in} \rangle$, $g_p=\langle \hat
p_{out} \rangle / \langle \hat p_{in} \rangle$. The adjusted gains
for the teleporters 1 and 2 are $g_{x1}=1.00\pm0.02$,
$g_{p1}=1.00\pm0.02$ and $g_{x2}=1.00\pm0.01$,
$g_{p2}=1.00\pm0.01$, respectively. For simplicity, these gains
are fixed throughout the experiment and treated as unity.

Bob receives Alice's measurement results ($x_u$, $p_v$) through
the classical channels and displaces his EPR beam ($\hat x_{\rm
B}$, $\hat p_{\rm B}$) accordingly, $ \hat x_{\rm B} \rightarrow
\hat x_{out}=\hat x_{\rm B} +\sqrt{2}x_u $ and $ \hat p_{\rm B}
\rightarrow \hat p_{out}=\hat p_{\rm B} +\sqrt{2}p_v $. In our
experiment, the displacement operations are realized via
electro-optical modulators (EOMs) and highly reflecting mirrors
(99/1 beam splitters). Bob modulates two beams by using amplitude
and phase modulators, corresponding to the displacement of $x$ and
$p$ quadratures, respectively. The modulated beams are combined
with Bob's mode ($\hat x_{\rm B}$, $\hat p_{\rm B}$) at the 99/1
beam splitters.

The teleported mode can be written as \cite{Takei05e}
\begin{align}
\hat x _{out}=\hat x_{in} -(\hat x_{\rm A}-\hat x_{\rm B}), \ \
\hat p _{out}=\hat p_{in} +(\hat p_{\rm A}+\hat p_{\rm B}).
\end{align}
Ideally, the EPR beams would have perfect correlations such that
$\hat x_{\rm A}-\hat x_{\rm B} \rightarrow 0$ and $\hat p_{\rm A}
+\hat p_{\rm B} \rightarrow 0$. Hence, the state of the output
mode, expressed by $\hat x_{out}$ and $\hat p_{out}$, would
coincide with that of the input mode, $\hat x_{in}$ and $\hat
p_{in}$.

\section{Experimental Results}

In the real experiment, the teleported state has some additional
noise due to the finite EPR correlations, i.e., $\Delta_{\rm
EPR}(x)\equiv\left \langle \left[ \Delta \left( \hat x_{\rm
A}-\hat x_{\rm B} \right) \right]^2\right \rangle \neq 0$ and
$\Delta_{\rm EPR}(p) \equiv \left \langle \left[ \Delta \left(
\hat p_{\rm A}+\hat p_{\rm B} \right) \right]^2\right \rangle \neq
0$. In the process of $n$ sequential quantum teleportations, this
excess noise is added $n$ times to the input state. Thus, the
variances of the output state are
\begin{align}
\langle ( \Delta \hat{x}_{out}^{(seq)} ) ^ 2 \rangle=\langle ( \Delta \hat{x}_{in} ) ^ 2 \rangle
+\sum_{i} \Delta_{{\rm EPR},i}(x),
\nonumber \\
\langle ( \Delta \hat{p}_{out}^{(seq)} ) ^ 2 \rangle=\langle ( \Delta \hat{p}_{in} ) ^ 2 \rangle
+\sum_{i} \Delta_{{\rm EPR},i}(p),
\end{align}
where $\Delta_{{\rm EPR},i}$ are the added noise terms of the
$i$-th teleporter.

   \begin{figure*}
     \begin{center}
        \includegraphics[width=1.0\linewidth]{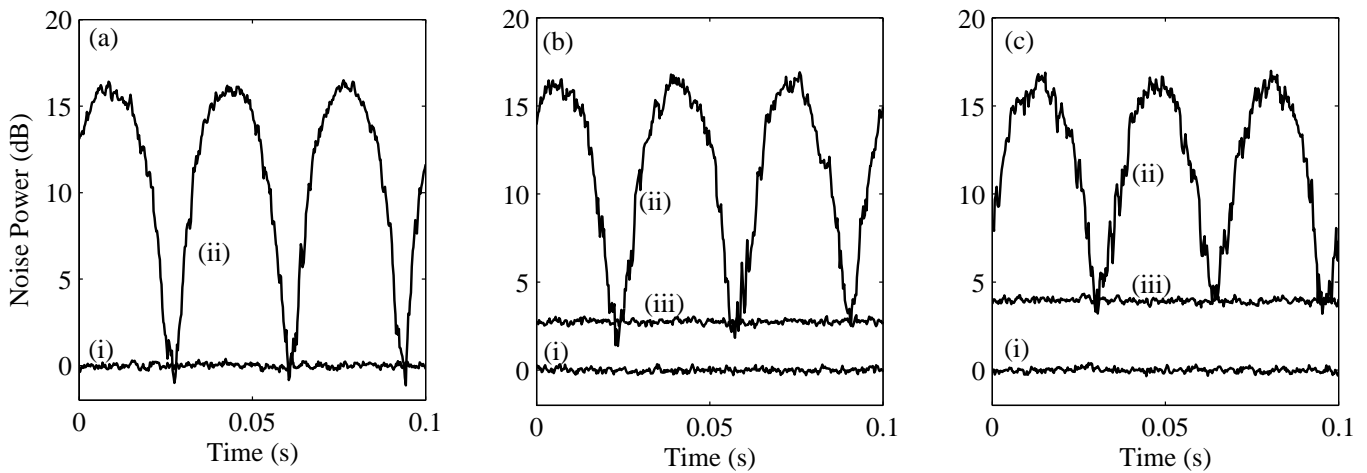}
        \caption{Measurement results of sequential teleportation for $x$ quadrature ( $p$ quadrature is not shown).
In all figures, traces (i) show vacuum noise level. (a) Input
coherent state. Trace (ii) shows the input state with phase
scanned. (b) Teleported states for $x$ quadrature. Trace (ii)
shows the teleported states for a coherent state input with a
phase of the input state scanned. Trace (iii) shows the teleported
states for a vacuum input of which noise level corresponds to the
variances of the output. The variance is $2.5 \pm 0.2$ dB (that of
$p$ is $2.8 \pm 0.2$ dB ). (c) Sequentially teleported state.
Trace (ii) shows  sequentially teleported states for a coherent
state input. Trace (iii) show the sequentially teleported states
for a vacuum input. The variance is $3.9 \pm 0.2$ dB (that of $p$
is $4.0 \pm 0.2$ dB ). The measurement frequency is 1 MHz,
resolution and video bandwidths are 30 kHz and 300 Hz,
respectively. All traces except for (ii) are averaged 30 times.}
     \end{center}
   \end{figure*}

Fig. 3 shows the measurement results of the two sequential quantum
teleportations. The outputs of the homodyne detection are measured
by a spectrum analyzer. The measurement frequency is 1 MHz. Fig. 3
(a) shows the input coherent state with the phase scanned. In our
experiment, an input coherent state is generated by modulating a
weak coherent beam at 1 MHz. Fig. 3 (b) shows the teleported
states for the $x$ quadrature ( the $p$ quadrature is not shown ).
The variances of the teleported state are $\langle (\Delta \hat
x_{out}^{(1)})^2\rangle= 2.5\pm0.2$ dB and $\langle (\Delta \hat
p_{out}^{(1)})^2\rangle= 2.8\pm0.2$ dB relative to the vacuum
noise level. Here superscript $(1)$ stands for the output of the
first teleporter. Fig. 3 (c) shows sequentially teleported states
for the $x$ quadrature ( the $p$ quadrature is not shown ). The
variances of the sequentially teleported state are $\langle
(\Delta \hat x_{out}^{(seq)})^2\rangle= 3.9\pm0.2$ dB and $\langle
(\Delta \hat p_{out}^{(seq)})^2\rangle= 4.0\pm0.2$ dB. Note that
the amplitudes of the teleported states are almost identical to
those of the input states, reassuring that the gains of the
teleporters are near unity.

We also evaluate the performance of the second teleporter
individually (not shown in Fig. 3). We teleport a coherent state
by using the second teleporter, and determine the variances of the
output. The measured variances of the output state are  $\langle
(\Delta \hat x_{out}^{(2)})^2\rangle= 2.3\pm0.2$ dB and $\langle
(\Delta \hat p_{out}^{(2)})^2\rangle= 2.2\pm0.2$ dB with respect
to the vacuum noise level. From Eq. (3), we can calculate the
variance of the sequentially teleported state from the added noise
of each teleporter. The calculated variances are $\langle (\Delta
\hat x_{out}^{(seq)})^2\rangle= 3.9$ dB and $\langle (\Delta \hat
p_{out}^{(seq)})^2\rangle= 4.1$ dB , which is in good agreement
with the experimental results. This confirms that our teleporters
maintain their fidelities even when combining them. In principle,
we can build a larger sequence of teleporters ( though at the
expense of a further decreasing total fidelity ).

To estimate the performance of a teleporter, we use the fidelity
$F=\langle \alpha |\hat{\rho}_{\rm out}|\alpha\rangle$ for
teleporting a coherent state with amplitude $\alpha$ yielding the
output state $\hat{\rho}_{\rm out}$. For coherent-state inputs
with unity gains, the fidelity can be written as \cite{Takei05e}
\begin{align}
F=\frac{2}{\sqrt{[1+4\langle(\Delta\hat{x}_{\rm out})^2\rangle]
[1+4\langle(\Delta\hat{p}_{\rm out})^2\rangle]}}.
\end{align}
The fidelity is calculated from the variances of the output
states. The variances of the coherent state input is $\langle (
\Delta \hat{x}_{in} ) ^ 2 \rangle=\langle ( \Delta \hat{p}_{in} )
^ 2 \rangle =1/4$, hence the fidelity can be determined by the
added noise of $\Delta_{\rm EPR}(x)$ and $\Delta_{\rm EPR}(p)$. We
calculate the fidelity from the measured variances using Eq. (4).
The performance of each teleporter is estimated for a
coherent-state input as  $F_{1}=0.70\pm0.02$ and
$F_{2}=0.75\pm0.02$ for teleporters 1 and 2, respectively. Note
that our teleporters exceed both the classical limit $F_{cl}=1/2$
\cite{Braunstein00,Braunstein01,Hammerer05} and the no-cloning
limit $F_{nc}=2/3$ \cite{Cerf00,Grosshans01}. We also calculate
the fidelity between the input and the sequentially teleported
state. The fidelity is $F^{(2)}=0.57\pm0.02$ which still exceeds
the classical limit $F_{cl}=1/2$ and verifies the successful
demonstration of two sequential quantum teleportations.

   \begin{figure*}
     \begin{center}
        \includegraphics[width=1.0\linewidth]{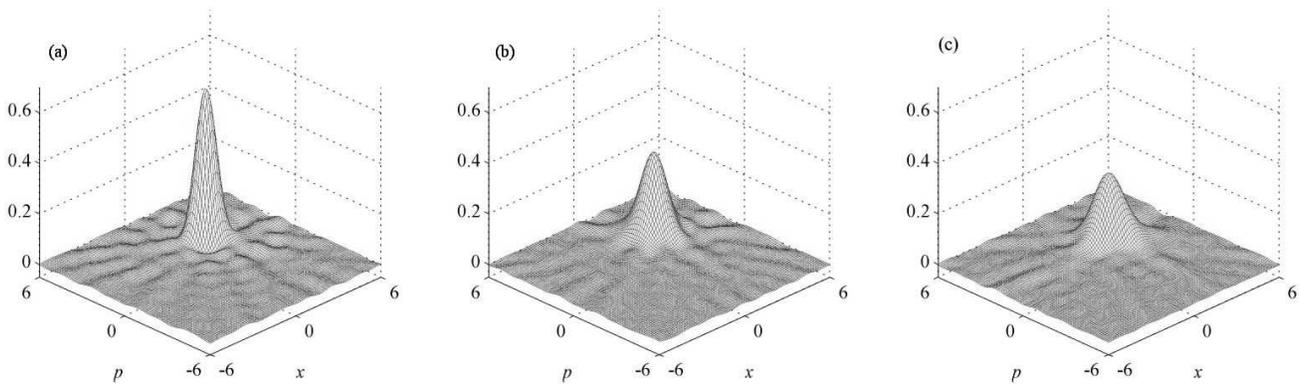}
        \caption{Wigner function reconstructed by using
optical homodyne tomography. (a) Input coherent state. (b)
Teleported state. (c) Sequentially teleported state. In these
measurement, we locked the phase of input coherent state $\sim$
45$^\circ$ from $x$ quadrature.}
     \end{center}
   \end{figure*}

Fig. 4 shows the Wigner functions reconstructed via optical
homodyne tomography technique \cite{Leonhardt_MQ(1997)}. The
Wigner function is a quasi-probability distribution defined by
   $W(x,p) = \frac{2}{ \pi }\int d\xi
                 \exp \left( -4i\xi p \right)
             \left\langle x+\xi \right|
                           \hat{\rho}
             \left| x-\xi \right\rangle$
\cite{Wigner32,Leonhardt_MQ(1997)}. In optical homodyne
tomography, we also use 1 MHz sidebands of the carrier beam
\cite{Breitenbach97}. The output of a homodyne detector is
high-pass filtered and then mixed with an electrical oscillator
signal of frequency 1 MHz, which is the same frequency as for the
creation of the input coherent state. The intermediate-frequency
output of the mixer is low-pass filtered (the bandwidth is 30 kHz)
and recorded in a PC with an analogue-to-digital converter (ADC).
The sampling rate of the ADC is set to 300 kHz and we measure
around 100,000 points with the local oscillator (LO) phase
scanned. To reconstruct the Wigner function, we use the inverse
Radon transformation \cite{Leonhardt_MQ(1997),Breitenbach97}. Fig.
4 (a) shows the input coherent state. Fig. 4 (b) shows the
teleported state and (c) shows the sequentially teleported state.
Note that here we lock the phase of the input coherent state to
$\sim 45^\circ$ with respect to the $x$ quadrature. Although the
teleported states have excess noise and turn into mixed states,
the input state is clearly reconstructed after the two rounds of
quantum teleportation.

\section{Summary}

In summary, we demonstrated two sequential quantum teleportations
of coherent states of light. The experimentally determined
fidelity for the sequentially teleported coherent states was well
beyond the classical limit. These results imply the possibility of
sequential quantum communication over several segments of a
channel using efficient Gaussian resources (combined with
non-Gaussian distillation techniques \cite{Browne03}). Moreover,
our experiment represents a first step towards sequentially
manipulating quantum information using entangled Gaussian resource
states. Future extensions of this work will include more general
measurement-based Gaussian transformations \cite{vanloock07} and
the use of non-Gaussian signal states as well as the addition of
non-Gaussian measurements in order to achieve universal quantum
information processing \cite{Menicucci06}.

\section{ACKNOWLEDGMENTS}
This work was partly supported by the MEXT of Japan.

\end{document}